\documentclass[dvipdfmx]{aastex631}
\UseRawInputEncoding

\usepackage[normalem]{ulem}
\usepackage{multirow}
\usepackage{graphicx}
\usepackage{float}
\shorttitle{The first detection of a protostellar outflow in the SMC}
\shortauthors{Tokuda et al.}
\graphicspath{{./}{figures/}}

\begin{document}

\title{The First Detection of a Protostellar CO Outflow in the Small Magellanic Cloud with ALMA}


\author[0000-0002-2062-1600]{Kazuki Tokuda}
\affiliation{Department of Earth and Planetary Sciences, Faculty of Science, Kyushu University, Nishi-ku, Fukuoka 819-0395, Japan}
\affiliation{National Astronomical Observatory of Japan, National Institutes of Natural Sciences, 2-21-1 Osawa, Mitaka, Tokyo 181-8588, Japan}
\affiliation{Department of Physics, Graduate School of Science, Osaka Metropolitan University, 1-1 Gakuen-cho, Naka-ku, Sakai, Osaka 599-8531, Japan}

\author[0000-0001-6149-1278]{Sarolta Zahorecz}
\affiliation{National Astronomical Observatory of Japan, National Institutes of Natural Sciences, 2-21-1 Osawa, Mitaka, Tokyo 181-8588, Japan}
\affiliation{Department of Physics, Graduate School of Science, Osaka Metropolitan University, 1-1 Gakuen-cho, Naka-ku, Sakai, Osaka 599-8531, Japan}

\author{Yuri Kunitoshi}
\affiliation{Department of Physics, Graduate School of Science, Osaka Metropolitan University, 1-1 Gakuen-cho, Naka-ku, Sakai, Osaka 599-8531, Japan}

\author{Kosuke Higashino}
\affiliation{Department of Physics, Graduate School of Science, Osaka Metropolitan University, 1-1 Gakuen-cho, Naka-ku, Sakai, Osaka 599-8531, Japan}

\author[0000-0002-6907-0926]{Kei E. I. Tanaka}
\affiliation{Center for Astrophysics and Space Astronomy, University of Colorado Boulder, Boulder, CO 80309, USA}
\affiliation{National Astronomical Observatory of Japan, National Institutes of Natural Sciences, 2-21-1 Osawa, Mitaka, Tokyo 181-8588, Japan}

\author[0000-0002-4098-8100]{Ayu Konishi}
\affiliation{Department of Physics, Graduate School of Science, Osaka Metropolitan University, 1-1 Gakuen-cho, Naka-ku, Sakai, Osaka 599-8531, Japan}

\author{Taisei Suzuki}
\affiliation{Department of Physics, Graduate School of Science, Osaka Metropolitan University, 1-1 Gakuen-cho, Naka-ku, Sakai, Osaka 599-8531, Japan}

\author{Naoya Kitano}
\affiliation{Department of Earth and Planetary Sciences, Faculty of Science, Kyushu University, Nishi-ku, Fukuoka 819-0395, Japan}

\author[0000-0002-8217-7509]{Naoto Harada}
\affiliation{Department of Earth and Planetary Sciences, Faculty of Sciences, Kyushu University, Nishi-ku, Fukuoka 819-0395, Japan}

\author[0000-0002-0095-3624]{Takashi Shimonishi}
\affiliation{Environmental Science Program, Faculty of Science, Niigata University, Ikarashi-ninocho 8050, Nishi-ku, Niigata, 950-2181, Japan}

\author[0000-0001-8901-7287]{Naslim Neelamkodan}
\affiliation{Department of Physics, United Arab Emirates University, Al-Ain, 15551, UAE}

\author[0000-0002-8966-9856]{Yasuo Fukui}
\affiliation{Department of Physics, Nagoya University, Furo-cho, Chikusa-ku, Nagoya 464-8601, Japan}

\author[0000-0001-7813-0380]{Akiko Kawamura}
\affiliation{National Astronomical Observatory of Japan, National Institutes of Natural Sciences, 2-21-1 Osawa, Mitaka, Tokyo 181-8588, Japan}

\author[0000-0001-7826-3837]{Toshikazu Onishi}
\affiliation{Department of Physics, Graduate School of Science, Osaka Metropolitan University, 1-1 Gakuen-cho, Naka-ku, Sakai, Osaka 599-8531, Japan}

\author[0000-0002-0963-0872]{Masahiro N. Machida}
\affiliation{Department of Earth and Planetary Sciences, Faculty of Science, Kyushu University, Nishi-ku, Fukuoka 819-0395, Japan}

\begin{abstract}

Protostellar outflows are one of the most outstanding features of star formation.
Observational studies over the last several decades have successfully demonstrated that outflows are ubiquitously associated with low- and high-mass protostars in the solar-metallicity Galactic conditions.
However, the environmental dependence of protostellar outflow properties is still poorly understood, particularly in the low-metallicity regime.
Here we report the first detection of a molecular outflow in the Small Magellanic Cloud with $0.2\:Z_\odot$, using Atacama Large Millimeter/submillimeter Array observations at a spatial resolution of $0.1{\rm\:pc}$ toward the massive protostar Y246. The bipolar outflow is nicely illustrated by high-velocity wings of CO(3--2) emission at $\gtrsim 15{\rm\:km\:s^{-1}}$. The evaluated properties of the outflow (momentum, mechanical force, etc.) are consistent with those of the Galactic counterparts. Our results suggest that the molecular outflows, i.e., the guidepost of the disk accretion at the small scale, might be universally associated with protostars across the metallicity range of $\sim0.2$--$1\:Z_\odot$.

\end{abstract}


\section{Introduction}\label{sec:intro}

Protostellar outflows are the most fundamental feedback in individual star formation. In the Galactic star-forming regions, outflows are ubiquitously observed as high-velocity wing features of molecular lines.
Since the first discovery of a molecular outflow from the low-mass protostar in L1551 IRS5 \citep{Snell_1980}, many outflows have been detected around both low- \citep[e.g.,][]{Fukui_1989,Arce_2007} and high-mass protostars \citep[e.g.,][]{Kurtz_2000,Beuther_2002,Zinnecker_Yorke_2007}. An outflow can be extended further than the subparsec scale, while it is launched from a much smaller region, i.e., an accretion disk of $\sim10$--$100{\rm\:au}$.
Magnetohydrodynamic (MHD) simulations with all protostellar masses have demonstrated that the outflow is magneto-centrifugally driven directly from the rotating circumstellar disk, unlike the classic jet-entrainment scenario \cite[e.g.,][]{Machida_2013, Matsushita_2017, Commercon_2022}.
High-resolution Atacama Large Millimeter/submillimeter Array (ALMA) observations also support this disk-wind mechanism around low- and high-mass protostars \citep[e.g.,][]{Bjerkeli_2016, Hirota_2017_2, Zhang_2018, Matsushita_2019}. 
However, previous outflow studies have mainly focused on the solar-neighborhood condition, and it is poorly understood whether such outflow-driving is universal even in non-Galactic environments.

It is important to understand star formation dynamics in metal-poor environments because metallicity increases with cosmic time. From a theoretical perspective, it is nontrivial if the disk-wind scenario also holds at lower metallicity. At lower metallicity and dust abundance, the efficient magnetic braking may prevent the formation of the rotating disk because the ionization rate increases and the dissipation rate of the magnetic field decreases \citep{Higuchi_2018}.
In addition, efficient dust cooling in low-metallicity environments could induce intense fragmentation of accretion disks \citep{Tanaka_2014, Matsukoba_2022}.
However, it is still unclear at what metallicity disk formation and outflow driving are inhibited due to the lack of long-term radiative MHD simulations of low-metallicity star formation. Therefore, there is a crucial need for observational constraints of star formation in low-metallicity environments, such as the outer Galaxy and the Magellanic Clouds.

Explorations of protostellar outflows in low-metallicity environments have made progress in recent years. An early single-dish CO study \citep{Brand_2007} detected a parsec-scale outflow in the cluster-forming region WB~89-789 at a galactocentric distance of $\sim$20\,kpc where the metallicity is four times lower than the solar value \citep[e.g.,][]{Fernadez_2017}. ALMA follow-up observations using various molecular lines, e.g., SiO and CCH, further determined the driving sources and revealed their high-velocity emission and cavity structures \citep{Shimonishi_2021}. 
\cite{Fukui_2015} discovered two extragalactic protostellar CO outflows from the high-mass young stellar objects (YSOs) with bolometric luminosities of $\sim$10$^{5}$\,$L_{\odot}$ in the N159W region of the Large Magellanic Cloud (LMC), which is a 0.5\,$Z_{\odot}$ environment. 
\cite{Tokuda_2019} resolved one of the outflows into several sources, whose properties, e.g., force, are roughly consistent with the Galactic relation \citep[e.g.,][]{Beuther_2002} expected from the protostellar bolometric luminosity and/or mass of the parental dense cores. \cite{Shimonishi_2016} subsequently reported an outflow from the hot core source ST11 in the LMC.  More recent ALMA observations toward a giant molecular cloud in M33, where the metallicity is similar to that of the LMC, found a promising protostellar outflow at a massive millimeter continuum source \citep{Tokuda_2020}. These discoveries strongly demonstrated the universality of the outflow activities associated with (high-mass) protostars within the local universe. However, any high-resolution CO observations in the Small Magellanic Cloud (SMC), which is in the lowest metallicity condition ($Z$ $\sim$0.2\,$Z_{\odot}$) among the molecular-gas-accessible Local Group of galaxies, have yet to discover molecular components of outflows from protostellar sources \citep{Muraoka_2017,Jameson_2018, Neelamkodan_2021,Tokuda_2021}. Because \cite{Ward_2017} reported shocked H$_2$ emission originating from outflows of massive YSOs in the SMC N81 and N88~A regions, the presence of outflows seems to be fairly likely. The molecular emission determination tracing most of the bulk of ejected material has been anticipated in the SMC to compare the physical properties of outflows among the Local Group sources and theoretical predictions, providing a key verification of the lunching mechanism of outflow in various conditions.

In this letter, we report on high-velocity ($\gtrsim$15\,km\,s$^{-1}$) wing emission in CO(3--2), which is an indication of molecular outflow from the embedded high-mass protostellar source Y246 in the SMC.

\section{Target description, observation, and data reduction} \label{sec:observe}

The Spitzer Space Telescope Surveying the Agents of Galaxy Evolution (SAGE)-SMC Legacy program \citep{Gordon_2011} has identified around a thousand intermediate- to high-mass YSO candidates in the SMC \citep{Sewilo_2013}, using color-magnitude diagrams, visual inspection of multi-wavelength images, and spectral energy distribution fitting with YSO models. \cite{Oliveira_2013} identified 33 spectroscopically confirmed intermediate- and high-mass YSOs. Among them, we have selected the most massive (brightest) six sources from the \cite{Oliveira_2013} YSO list as the ALMA follow-up targets. Two of them (\#17 and 18) are likely the most embedded stage because they show both ice and silicate absorption features. The others (\#03, 13, 28, and 33) have warmer envelopes since they show no ice absorption and/or polycyclic aromatic hydrocarbons and optical fine-structure emission. 

For our observations, the ALMA 12\,m array's configuration was set to C43-4, using the Band~7 receiver (275-373\,GHz). We used three basebands, whose bandwidth and frequency resolution were 937.5\,MHz and 0.564\,MHz, respectively. The central frequencies were 356.76, 344.20, and 345.80\,GHz. The remaining baseband was used for only continuum observations, with a bandwidth of 1875\,MHz centered at 355.00\,GHz.
This letter focuses on the analysis of an outflow tracer CO(3--2) and a dense-gas tracer SO($N,J$ = 8, 8--7, 7) along with the continuum emission to investigate gas properties around the protostellar sources. We will present other targets and their hot core study in a separate paper (S., Zahorecz et al. in preparation.).
We performed the data reduction and imaging using the Common Astronomy Software Application package, ver~5.6.1-8. 
We used the \texttt{tclean} task with the multiscale deconvolver. We applied the Briggs weighting with a robust parameter of 0.5. We made velocity cubes of CO(3--2) and SO($N,J$ = 8,8--7,7) with a velocity bin of 0.5\,km\,s$^{-1}$. The resultant beam size and sensitivity of the molecular lines are 0$\farcs$45 $\times$ 0$\farcs$34 and 3\,mJy\,beam$^{-1}$ (= 0.2\,K), respectively. At the distance of the SMC ($\sim$62\,kpc; \citealt{Graczyk_2020}), the spatial resolution is $\sim$0.1\,pc. We imaged the continuum data with an aggregate bandwidth of $\sim$4.5\,GHz to avoid strong CO(3--2) emission. The beam size and sensitivity are 0$\farcs$41 $\times$ 0$\farcs$30 and $\sim$0.07\,mJy\,beam$^{-1}$, respectively.

Our visual inspection of the CO(3--2) emission found that Y246 \citep{Sewilo_2013}, which corresponds to \#18 in the \cite{Oliveira_2013} catalog, with a bolometric luminosity of $\sim$2.8 $\times$10$^{4}$\,$L_{\odot}$ is the most promising candidate for a molecular outflow (Sections~\ref{ssR:outflow} and \ref{ssD2}). 
The observed central coordinate of Y246 was ($\alpha_{J2000.0}$, $\delta_{J2000.0}$)=(0$^{\rm h}$54$^{\rm m}$03\fs36, -73\arcdeg19\arcmin38\farcs4), which corresponds to the coordinate determined by the early Spitzer-based identification \citep{Sewilo_2013,Oliveira_2013}.


\section{Results} \label{sec:results}

\subsection{Continuum emission and molecular outflow} \label{ssR:outflow}

\begin{figure*}[htbp]
\centering
\includegraphics[width=170mm]{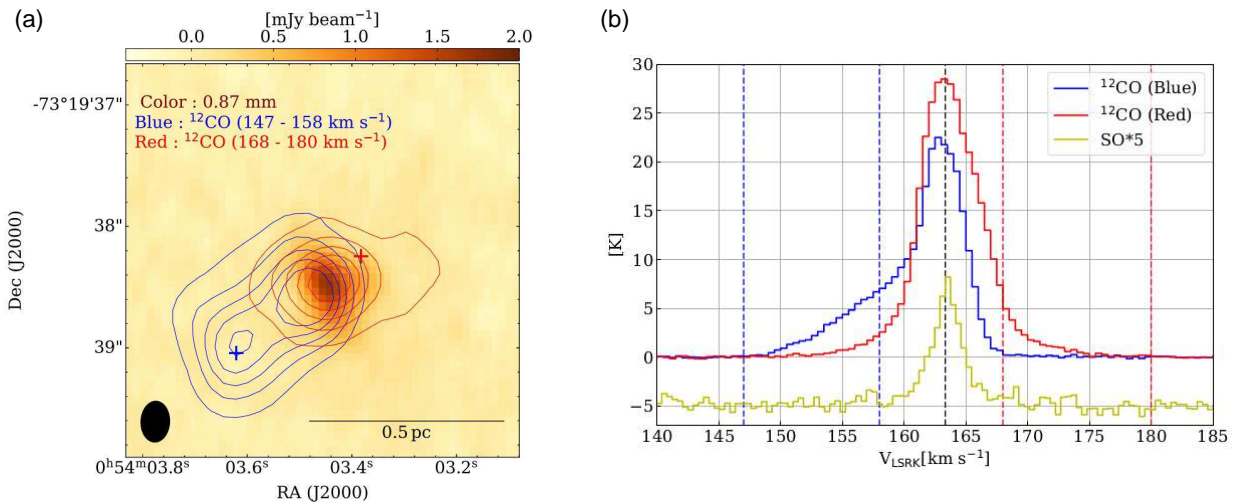}
\caption{0.87\,mm continuum and CO(3--2) high-velocity emission toward the Spitzer source. (a) Color-scale image shows the 0.87\,mm continuum emission. The ellipse at the lower-left corner shows the beam size, 0$\farcs$41 $\times$ 0$\farcs$30. Blue and red contours show the CO(3--2) high-velocity wing emission with integrated velocity ranges of 147--158\,km\,s$^{-1}$ and 168--180\,km\,s$^{-1}$, respectively (see also dotted lines in panel (b)). The lowest and subsequent contour steps are 6 and 10\,K\,km\,s$^{-1}$, respectively. Note that the rms noise level is $\sim$1.3\,K\,km\,s$^{-1}$. (b) The blue and red profiles show the CO blueshifted and redshifted high-velocity emission, respectively, extracted from a 0\farcs45 radius centered at red/blue crosses in panel (a). The yellow profile shows the SO spectrum at the 0.87\,mm continuum peak. Note that the intensity is multiplied by a factor of 5 and then is shifted $-$5\,K for visualization purposes. The black dotted lines indicate the systemic velocity of 163\,km\,s$^{-1}$ determined by the Gaussian fitting. 
\label{fig:outflow}}
\end{figure*}

The color-scale image in Figure~\ref{fig:outflow}a shows the 0.87\,mm continuum emission toward the Spitzer source Y246. We performed 2D Gaussian fitting to the source emission, and obtained the following parameters: a central coordinate of ($\alpha_{J2000.0}$, $\delta_{J2000.0}$)=(0$^{\rm h}$54$^{\rm m}$03\fs439, -73\arcdeg19\arcmin38\farcs528), a beam-deconvolved size of 0\farcs69 $\times$ 0\farcs48 (=0.21\,pc $\times$ 0.14\,pc) with a position angle of 27$^{\circ}$. 
The difference between the source position determined by Spitzer and the dust peak is $\sim$0\farcs4, which is not a significant offset. The distribution of the continuum emission strongly suggests that it is emitted from the dense envelope gas, mainly as thermal dust emission, associated with the protostellar system. 
The total 0.87\,mm flux ($F_{\rm 0.87\,mm}$) is 6.9\,mJy. Assuming a uniform dust temperature ($T_{\rm d}$) of 20--30\,K \citep{Takekoshi_2017}, a gas-to-dust ratio of 1000  \citep[e.g.,][]{Roman-Duval_2014}, and a dust opacity per dust mass of 2\,cm$^{2}$\,g$^{-1}$ \citep{Ossenkopf_1994}, the $F_{\rm 0.87\,mm}$ of Y246 can be converted into 0.7 $\times$10$^{3}$\,$M_{\odot}$ (for $T_{\rm d}$ = 30\,K) and to 1.2 $\times$10$^{3}$\,$M_{\odot}$ (for $T_{\rm d}$ = 20\,K). The velocity width in FWHM of the SO profile at the 0.87\,mm peak is 3.4\,km\,s$^{-1}$. With a radius of 0.17\,pc, the estimated virial mass is $\sim$10$^{3}$\,$M_{\odot}$ using the method in \cite{MacLaren_1988} assuming a uniform density distribution. Since the virial mass is about the same or smaller than the dust-based mass, it is reasonable to assume that this high-density clump is gravitationally bound. The amount of surrounding gas is large enough to be consistent with the fact that the protostar is in a young evolutionary stage (see Section~\ref{sec:observe}).

The CO(3--2) data enables us to probe molecular outflow around the Spitzer source. We visually inspected the presence/absence of the outflow wing emission, whose maximum velocity is apart by more than $\sim$10\,km\,s$^{-1}$ from the systemic velocity of 163\,km\,s$^{-1}$ determined by the Gaussian fitting of the SO profile. The blueshifted and redshifted CO wing emission in Figure~\ref{fig:outflow}b shows a maximum velocity ($v_{\rm max}$) of $\sim$16\,km\,s$^{-1}$ and $\sim$17\,km\,s$^{-1}$, respectively. We superimposed the high-velocity emission integrated over the velocity range of 147--158\,km\,s$^{-1}$ (blueshifted) and 168--180\,km\,s$^{-1}$ (redshifted) on the 0.87\,mm continuum map (Figure~\ref{fig:outflow}a). 
Although the blueshifted profile extends to lower velocity, $\sim$160\,km\,s$^{-1}$, we did not integrate the velocity range due to strong contamination from around the systemic velocity components. The blue and redshifted components roughly compose a northwest to southeast elongation centered at the continuum peak, which corresponds to the position of the embedded source. 
These spatial distributions and velocity features are quite consistent with known outflow sources in the Galaxy and LMC (see Section~\ref{sec:intro} and Section~\ref{ssD1}) and nicely illustrate the protostellar, molecular outflow from the embedded source.
Note that no other high-velocity components similar to those in the 0.87 mm source were detected within the observation field of the $\sim$25$\arcsec$ diameter.

We estimated the physical properties of the outflow lobes. The projected length of the lobes ($R_{\rm lobe}$) is $\sim$0.3--0.4\,pc, based on the distance between the edges of the outflow lobes and the continuum peak. The dynamical time ($t_{\rm d}$ = $R_{\rm lobe}$/$v_{\rm max}$) of the lobes is 2$\times$10$^{4}$\,yr (without inclination angle correction), indicating the protostellar system is in a young evolutionary stage, consistent with the spectroscopic analysis by \cite{Oliveira_2013}. The CO(3--2) luminosities of the blue and redshifted lobes are $\sim$4\,K\,km\,s$^{-1}$\,pc$^{2}$ and $\sim$2\,K\,km\,s$^{-1}$\,pc$^{2}$, respectively. Although the CO-to-H$_2$ conversion factor is not well constrained in the SMC-like low-metallicity environment, we applied the recently derived value of 7.5 $\times$10$^{20}$\,cm$^{-2}$\,(K\,km\,s$^{-1}$)$^{-1}$ \citep{Muraoka_2017} to estimate the flow masses ($M_{\rm flow}$) as $\sim$70\,$M_{\odot}$ (for the blue lobe) and $\sim$30\,$M_{\odot}$ (for the red lobe).  




\section{Discussion} \label{sec:discuss}

\subsection{The importance of the first discovery of an SMC protostellar outflow}\label{ssD1}

We compare the physical properties of the newly discovered SMC protostellar outflow with those in the Galactic and LMC studies based on an order estimation. The mass ejection rate ($M_{\rm flow}$/$t_{\rm d}$) is $\sim$10$^{-3}$\,$M_{\odot}$\,yr$^{-1}$, which is consistent with the fact that the protostellar source is in a young evolutionary stage of high-mass star formation. The outflow momentum, $P_{\rm flow}$ = $M_{\rm flow} v_{\rm max}$, is $\sim$10$^{3}$\,$M_{\odot}$\,km\,s$^{-1}$, and the mechanical force, $F_{\rm flow}$ = $P_{\rm flow}/t_{\rm d}$, is $\sim$10$^{-2}$\,$M_{\odot}$\,km\,s$^{-1}$\,yr$^{-1}$. These outflow properties are roughly consistent with those seen in the Galactic protostars (see Figure~5 in \citealt{Beuther_2002} and Figure~7 in \citealt{Maud_2015}) and in the solar-metallicity MHD simulations (see Figures~5 and 6 of \citealt{Matsushita_2018}), at the source luminosity of $\sim$10$^{4}$\,$L_{\odot}$. \cite{Tokuda_2019} also confirmed this relation in a luminous YSO in the LMC N159W-South region. Thus, we conclude that the outflow properties may be universal across the metallicity range of 0.2--1\,$Z_{\odot}$, based on the current observational knowledge of the Galaxy, LMC, and SMC.

We would like to note that, as mentioned in Section~\ref{sec:intro}, it is nontrivial whether disks form and drive outflows in low-metallicity environments from a theoretical perspective. The detection of the molecular outflow in Y246 suggests that an accretion disk exists under a strong magnetic field in the unresolved innermost region, even in the SMC with $0.2\:Z_\odot$. Galactic observations indicate that magnetic fluxes and angular momenta in prestellar cores are several orders of magnitude larger than those of protostars \citep[e.g.,][]{Goodman_1993,Crutcher_1999}, known as the magnetic-flux and angular-momentum problems. Theoretical studies on the Galactic star formation have investigated the mechanisms to remove the large amounts of angular momentum and magnetic flux in star formation. An excess angular momentum is transported by protostellar outflow driven by the outer edge of the disk, while nonideal MHD dissipation processes can remove magnetic flux from a star-forming core and promote the formation of a rotationally supported disk \citep[e.g.,][]{Dapp_2010}. In addition, the dissipation of magnetic fields is closely related to disk fragmentation, and the formation of binary or multiple stellar systems \cite[e.g.,][]{Machida_2008}. As the dissipation rate of magnetic fields is suggested to be controlled by the metallicity, the outflow intensity and binary frequency should also depend on the metallicity \citep[e.g.,][]{Susa_2015}. Moreover, the dust abundance affects radiative heating and cooling rates, changing the gravitational stability of accretion disks \citep[e.g.,][]{Matsukoba_2022}. 
Therefore, it is crucial to understand the metallicity dependence of radiative MHD dynamics in star formation. However, so far theorists have primarily assumed solar metallicity, and only a few studies have focused on the outflow driving and disk formation in low-metallicity environments \citep{Higuchi_2018, Tanaka_2018}. Thus, to comprehensively understand the star formation process across various environments, we must shed light on low-metallicity star formation in future theoretical studies.

The high-mass protostar Y246 is currently only one example in the SMC, but it will be an important benchmark for future theoretical and observational studies of low-metallicity star formation. To better understand star formation dynamics in the low-metallicity SMC environment, we will need to examine the physical properties of the surrounding dense envelope, e.g., specific angular momentum and infall signature through higher-resolution follow-ups.

\subsection{Are molecular outflows rare in the SMC?}\label{ssD2}

It has been for seven years since the first extragalactic protostellar outflow detection in the LMC \citep{Fukui_2015}. We here discuss why such a discovery has been delayed in the SMC. We mainly point out three possibilities: the sample size, the tracer effect, and the scarcity in nature.

As for the sample size, the known YSO targets with the spectroscopic confirmation in the LMC and SMC are 277 \citep{Seale_2009} and 33 \citep{Oliveira_2013}, respectively. 
The difference of one order of magnitude is proportional to the sizes of the galaxies and their CO-detectable regions \citep[e.g.,][]{Fukui_2010}. The large sample size of luminous ($\gtrsim$10$^{4}$\,$L_{\odot}$) YSOs in the LMC naturally explains the high encounter probability, as higher YSO luminosities generally tend to show energetic, detectable outflows.

\begin{figure*}[htbp]
\centering
\includegraphics[width=110mm]{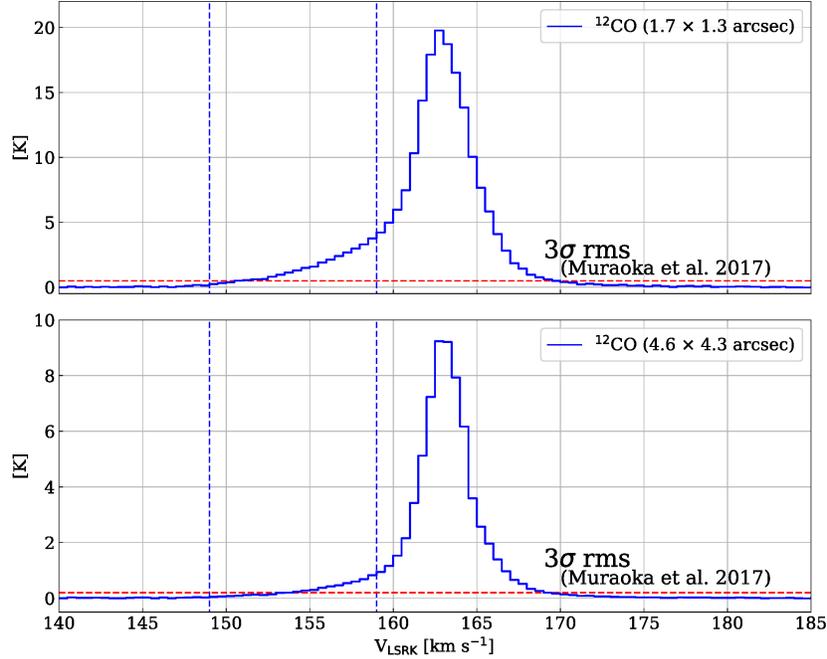}
\caption{
Upper and lower profiles show the smoothed CO(3--2) spectra at angular resolutions of 1\farcs7 $\times$ 1\farcs3 and 4\farcs6 $\times$ 4\farcs3, respectively, in Y246. The extracted position of the profiles is the same as that of the blue cross in Figure~\ref{fig:outflow}. The horizontal red lines represent the 3$\sigma$ rms noise levels of CO(2--1) (upper panel) and CO(3--2) (lower panel) observations toward N83C by \cite{Muraoka_2017}. The blue vertical lines are the same as those in Figure~\ref{fig:outflow}b.  
\label{fig:smooth_spect}}
\end{figure*}

Previous CO observations mainly studied $J$ = 1--0 and 2--1 transitions in the SMC's molecular clouds, i.e., the tracer effect may give a reason for the outflow nondetection results. For example, the CO(2--1) observations by \cite{Muraoka_2017} could not detect the molecular outflow from the high-mass protostar in the SMC N83C, although the target is as bright as $\sim$10$^{5}$\,$L_{\odot}$ and the observational setting was similar to the LMC studies \citep[e.g.,][]{Fukui_2015}.
We spatially smoothed our Y246 CO(3--2) data set into an angular resolution of 1\farcs7 $\times$ 1\farcs3, which is the same as the \cite{Muraoka_2017} observation on N83C (see Figure~\ref{fig:smooth_spect} upper panel). The resultant brightness temperature of the wing emission is $\gtrsim$1\,K, which can be detected with $>$3$\sigma$ at the \cite{Muraoka_2017} sensitivity assuming a CO(3--2)/CO(2--1) ratio of $\sim$1. The nondetection in N83C was probably attributed to a weaker low-$J$ CO emission in high-velocity outflow components. 
Brightness temperatures of higher-$J$ transitions become higher than those in lower $J$ ones in the case of being optically thin and high temperature ($>$20--30\,K) under the local thermodynamical equilibrium conditions.
Such environments are rarely seen in typical CO-detectable molecular clouds, but some local regions show stronger CO(3--2) and higher $J$ emission than lower-$J$ CO lines, e.g., at a cloud edge \citep{Tachihara_2012} and protostellar outflows \citep[e.g.,][]{Shimajiri_2008,Lefloch_2015}.
This behavior may be further enhanced in the SMC because CO lines trace denser and warmer gas in lower-metallicity environments \cite[e.g.,][]{Muraoka_2017,Tokuda_2021} than in typical giant molecular clouds in the Galaxy \cite[e.g.,][]{Nishimura_2015}. We thus speculate that CO(3--2) worked favorably as an outflow tracer in the SMC. 
We note that the available CO(3--2) data in N83C \citep{Muraoka_2017} has a further coarse angular resolution of $\sim$4\arcsec, which did not allow us to detect weak, compact emissions due to the beam-dilution effect. Based on a smoothed spectra of Y246 (Figure~\ref{fig:smooth_spect} lower panel), it becomes more difficult to detect the high-velocity emission with a relative velocity of $\gtrsim$10\,km\,s$^{-1}$. 

As mentioned in the previous paragraph, the behavior of CO as a dense-gas tracer may explain the scarcity of the outflow detection in the SMC. The lobe size of the Y246 outflow (0.3--0.4\,pc) is considerably smaller than the parsec-scale CO outflow in the outer Galactic region \citep{Fernadez_2017}. The time and spatial scales with observable CO outflows may be limited in the SMC, which may further reduce the probability of detection. However, our preliminary analysis found one or two more wing feature candidates (\#03, and 33) in the current ALMA sample from the \cite{Oliveira_2013} catalog, and thus it is premature to conclude that CO outflows are rare in the SMC.
Further multiobject with multitransition observations will be anticipated to obtain a convincing population of SMC outflows and to investigate their similarities/differences compared to those in the Galaxy and the LMC.

\section{Summary} \label{sec:summary}

We performed molecular-line and continuum observations toward the Spitzerーidentified YSO Y246 in the SMC using ALMA with the Band~7 receiver at a spatial resolution of $\sim$0.1\,pc. The 0.87\,mm continuum and dense-gas tracer identified a dense clump whose size and mass are $\sim$0.1\,pc and more than a few hundred solar masses, respectively, at the position of the Spitzer source. The CO(3--2) emission around the protostellar source clearly shows high-velocity ($\gtrsim$15\,km\,s$^{-1}$) bipolar features. This is the first discovery of a CO protostellar outflow in the SMC, which is the lowest metallicity ($Z$ $\sim$0.2\,$Z_{\odot}$) galaxy among the known CO-accessible Local Group of galaxies. 
Our new findings suggest that the (high-mass) star formation process driving a molecular outflow may be common throughout various environments in the local universe with a metallicity of 0.2--1\,$Z_{\odot}$.

\begin{acknowledgments}

We would like to thank the anonymous referee for useful comments that improved the manuscript. This paper makes use of the following ALMA data: ADS/JAO. ALMA\#2019.1.00534.S. ALMA is a partnership of ESO (representing its member states), the NSF (USA), and NINS (Japan), together with the NRC (Canada), MOST, and ASIAA (Taiwan), and KASI (Republic of Korea), in cooperation with the Republic of Chile. The Joint ALMA Observatory is operated by the ESO, AUI/NRAO, and NAOJ. This work was supported by a NAOJ ALMA Scientific Research grant Nos. 2022-22B, Grants-in-Aid for Scientific Research (KAKENHI) of Japan Society for the Promotion of Science (JSPS; grant No. JP18H05440, JP19K14760, JP21H00049, JP21H00058, JP21H01145, and JP21K13962). 
N. N. acknowledges a United Arab Emirates University grant No. UPAR G00003479.

\end{acknowledgments}

\end{document}